\documentclass[useAMS,usenatbib]{mn2e}
\usepackage{citesort,epsfig,psfrag,array}
\usepackage{amsmath,amsfonts,amssymb}
\usepackage{natbib}
\usepackage{graphicx}
\usepackage{mathptmx}
\usepackage{dcolumn}
\usepackage{subfig}
\usepackage{journals}
\newcolumntype{d}{D{.}{.}{-1}}

\graphicspath{{./}}
\newcounter{todo}
\renewcommand\thetodo{\Alph{todo}}
\def\todo#1{\addtocounter{todo}{1}[[\thetodo: #1]]\strut\vadjust{%
\kern-\dp\strutbox{\vtop to \dp\strutbox{\baselineskip\dp\strutbox\vss\rlap{%
\hskip\hsize\ \rm{$\leftarrow$\thetodo}}\null}}}}
\def\note#1{\strut\vadjust{\kern-\dp\strutbox{\vtop to \dp\strutbox{%
\baselineskip\dp\strutbox\vss\rlap{\hskip\hsize\ {\tiny\rm #1}}\null}}}}
\newcommand{\tablefont}{\fontsize{6}{8}\selectfont}
\title[SZ observations of two \textit{Planck} clusters with AMI]
{Further Sunyaev--Zel'dovich observations of two \textit{Planck} ERCSC clusters with the Arcminute Microkelvin Imager\thanks{We request that any reference to this paper cites ``AMI Consortium: Hurley-Walker et al. 2011"}}

\label{firstpage}

\author[Hurley-Walker et~al.]
{AMI Consortium:
Natasha~Hurley-Walker$^{1}$\thanks{Issuing author; E-mail: nh313@mrao.cam.ac.uk},
Michael~L.~Brown$^{1,2}$, Matthew~L.~Davies$^{1}$,\newauthor
Farhan~Feroz$^{1}$, Thomas M. O. Franzen$^{3}$, Keith~Grainge$^{1,2}$, Michael~P.~Hobson$^{1}$ \newauthor
Anthony~Lasenby$^{1,2}$, Malak~Olamaie$^{1}$, Guy~Pooley$^{1}$, Carmen~Rodr\'{\i}guez-Gonz\'alvez$^{1}$, \newauthor
Richard~D.~E.~Saunders$^{1,2}$, Michel~P.~Schammel$^{1}$, Anna~M.~M.~Scaife$^{4}$, Paul~F.~Scott$^{1}$, \newauthor
Timothy Shimwell$^{1}$, David~Titterington$^{1}$ and Elizabeth~Waldram$^{1}$\\
 $^1$ Astrophysics Group, Cavendish Laboratory,
      19 J.~J.~Thomson Avenue, Cambridge CB3 0HE \\
 $^2$ Kavli Institute for Cosmology, Cambridge, Madingley Road, Cambridge CB3 0HA\\
 $^3$ CSIRO Astronomy \& Space Science, PO Box 76, Epping, NSW 1710, Australia\\
 $^4$ Dublin Institute for Advanced Studies, 31 Fitzwilliam Place,
      Dublin 2, Ireland}

\date{Accepted ---; received ---; in original form \today}
\pagerange{\pageref{firstpage}--\pageref{lastpage}}
\pubyear{2011}

\begin{document}
\maketitle

\begin{abstract}
We present follow-up observations of two galaxy clusters detected blindly via the Sunyaev-Zel'dovich (SZ) effect
and released in the \textit{Planck} Early Release Compact Source Catalogue. 
We use the Arcminute Microkelvin Imager, a dual-array 14--18\,GHz radio interferometer. After radio source subtraction, we find a SZ decrement of integrated flux density $-1.08\pm0.10$\,mJy
toward PLCKESZ\,G121.11+57.01, and improve the position measurement of the cluster, finding the centre to be
RA~12$^{\mathrm{h}}$~59$^{\mathrm{m}}$~36.4$^{\mathrm{s}}$, Dec~+60$^{\circ}$04\arcmin~46\farcs8, to an accuracy of 20\arcsec.
The region of PLCKESZ\,G115.71+17.52 contains strong extended emission, so we are
unable to confirm the presence of this cluster via the SZ effect.
\end{abstract}

\begin{keywords}
   cosmology: observations -- radiation mechanisms: non-thermal -- cosmic microwave background
-- galaxies: clusters -- Sunyaev-Zel'dovich  
-- galaxies: clusters: individual (PLCKESZ\,G115.71+17.52, PLCKESZ\,G121.11+57.01)
\end{keywords}

\section{Introduction}

The \textit{Planck} satellite is an instrument designed to measure the full
Cosmic Microwave Background (CMB) sky to the highest sensitivity yet achieved. 
Part of its remit is a blind survey of galaxy clusters via their Sunyaev-Zel'dovich effects (SZ; \citealt{sz70});
this occurs when CMB photons are inverse-Compton scattered by hot gas
in galaxy clusters. The resulting shift in the blackbody spectrum of the CMB
produces a decrement at radio frequencies below the null of 217\,GHz and a positive signal above
this frequency (see e.g.~\cite{birkinshaw99} and \cite{carlstrom02} for reviews).

The Arcminute Microkelvin Imager (AMI) is a dual-array radio interferometer operating at 14--18\,GHz;
part of its scientific programme 
is making observations of hundreds of known galaxy clusters to measure their gas masses and structures.
Along with its high sensitivity, its resolution of $\simeq2\farcm5$ allows the measurement of 
the extent and shape of galaxy clusters above a redshift of $\simeq0.1$ (below which they become resolved out).

In 2011, the \textit{Planck} Early Release Compact Source Catalogue (ERCSC; \citealt{PlanckERCSC})
was made publicly available; it includes a sample of 189 galaxy clusters detected via their SZ effects
at signal-to-noise (S/N)\,$>6$ \citep{PlanckESZ}. 
AMI was used to follow up the two previously unconfirmed clusters lying in the northern sky:
PLCKESZ\,G115.71+17.52 (S/N$=6.78$) and PLCKESZ\,G121.11+57.01 (S/N$=6.66$), 
AMI observing details of which can be found in Tab.~\ref{tab:obslist}.

\begin{table*}
\begin{center}
\begin{tabular}{lccccccc}
\hline
Name & RA & Dec & SA observing & SA map noise / & LA observing& LA 19-pt map& LA 61-pt map\tabularnewline
    & (J2000) & (J2000) & time / hrs & $\mu$Jy\,beam$^{-1}$ & time / hrs & noise/ $\mu$Jy\,beam$^{-1}$ & noise/ $\mu$Jy\,beam$^{-1}$\tabularnewline
\hline
PLCKESZ\,G115.71+17.52 & 22 26 24.9 & +78 18 15.8 & 50 & 78 & 77 & 55 & 200 \tabularnewline
PLCKESZ\,G121.11+57.01 & 12 59 23.8 & +60 05 24.8 & 40 & 81 & 67 & 65 & 250 \tabularnewline
\hline
\end{tabular}

\caption{Statistics of the AMI observations of the two candidates.
\label{tab:obslist}}

\end{center}
\end{table*}

%Throughout, we assume a concordance $\Lambda$CDM cosmology, with $\Omega_{\mathrm{m}}=0.3$,
%$\Omega_{\Lambda}=0.7$, $H_{0}=100$km\,s$^{-1}$Mpc$^{-1}$ (and thus $h=1.0$). All
%coordinates are J2000 epoch.

\section{AMI SZ observations}\label{sec:SZ}

AMI comprises two aperture synthesis arrays located at the Mullard Radio
Astronomy Observatory, Lord's Bridge, Cambridge, UK. The AMI Small Array (SA) consists
of ten 3.7-m diameter equatorially-mounted dishes with a baseline range of
$\simeq 5$--20\,m, while the AMI Large Array (LA) has eight 12.8-m-diameter dishes
with a baseline range of $\simeq 20$--100\,m. Both arrays observe I+Q in the band 14--18\,GHz, each with
system temperatures of about 25\,K.

The backends are analogue Fourier
transform spectrometers, from which the complex signals in each of eight
channels of 750-MHz bandwidth are synthesised, and the signals in the synthesised channels
are correlated at the $\simeq 10$~per~cent level. In practice, the
two lowest-frequency channels are generally not used due to a poor correlator response in
this frequency range and interference from geostationary satellites. Further telescope details are given in
\cite{2008MNRAS.391.1545Z}.

The AMI data reduction was performed using our in-house reduction software \textsc{reduce}.
This is used to apply path-compensator and path-delay corrections, to flag interference,
shadowing and hardware errors, to apply phase and amplitude calibrations, make a correction for
the changing intervening air mass and to
Fourier transform the correlator data to synthesise the frequency channels,
before outputting to disk in $uv$-\textsc{fits} format suitable for imaging in the
Astronomical Image Processing System (\textsc{aips}; \citealt{aips}).
Flux calibration was performed using short observations of 3C48 and 3C286 near
the beginning and end of each run, with assumed I+Q flux densities for these sources in
the AMI channels consistent with \citet{1977A+A....61...99B} (see Tab.~\ref{tab:Fluxes-of-3C286}).
As \citeauthor{1977A+A....61...99B} measure I and AMI measures I+Q, these flux densities
include corrections for the polarization of the sources.
The phase calibrators, J0017+8135, J1302+5748 and J2133+8239, were chosen from the
Jodrell Bank VLA Survey
(JVAS; \citealt{1992MNRAS.254..655P}) on the basis of proximity ($\leq5^\circ$)
and flux density ($\geq0.7$\,Jy at 15\,GHz).
After phase calibration, the phase of AMI over one hour is generally
stable to $5^{\circ}$ for channels 4--7, and to $10^{\circ}$ for channels 3 and 8.
The overall consistency of the
flux-density scale is estimated to be better than five~per~cent.

\begin{table}
\centering
\begin{tabular}{cccccc}\hline
 Channel & $\nu$/GHz & $S^{{\rm {3C286}}}$/Jy & $S^{{\rm {3C48}}}$/Jy &
 FWHM$_{\mathrm{LA}}$/\arcmin \\ \hline
 3 & 13.9 & 3.74 & 1.89 & 6.08\\
 4 & 14.6 & 3.60 & 1.78 & 5.89\\
 5 & 15.3 & 3.47 & 1.68 & 5.70\\
 6 & 16.1 & 3.35 & 1.60 & 5.53\\
 7 & 16.9 & 3.24 & 1.52 & 5.39\\
 8 & 17.6 & 3.14 & 1.45 & 5.25\\ \hline
\end{tabular}
\caption{Assumed I+Q flux densities of 3C286 and 3C48 over the commonly-used AMI band, and the
LA primary beam full-width-half-maxima (FWHM) for each channel.  \label{tab:Fluxes-of-3C286}}
\end{table}

\begin{figure}
\begin{center}
\subfloat[PLCKESZ\,G115.71+17.52\label{fig:PLJ2226_LA}]{\includegraphics[width=8cm,clip=]{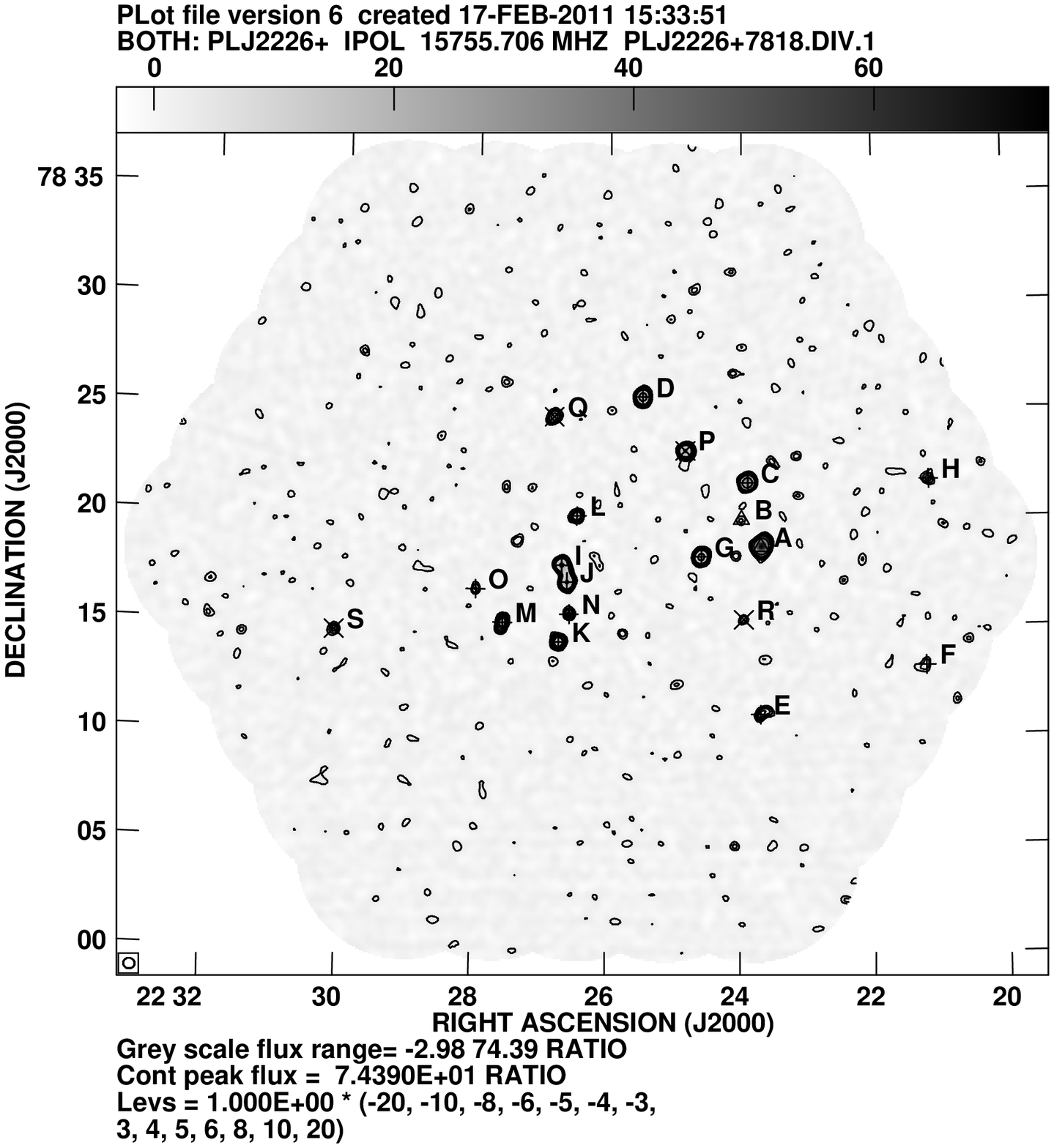}}\quad
\subfloat[PLCKESZ\,G121.11+57.01\label{fig:PLJ1259_LA}]{\includegraphics[width=8cm,clip=]{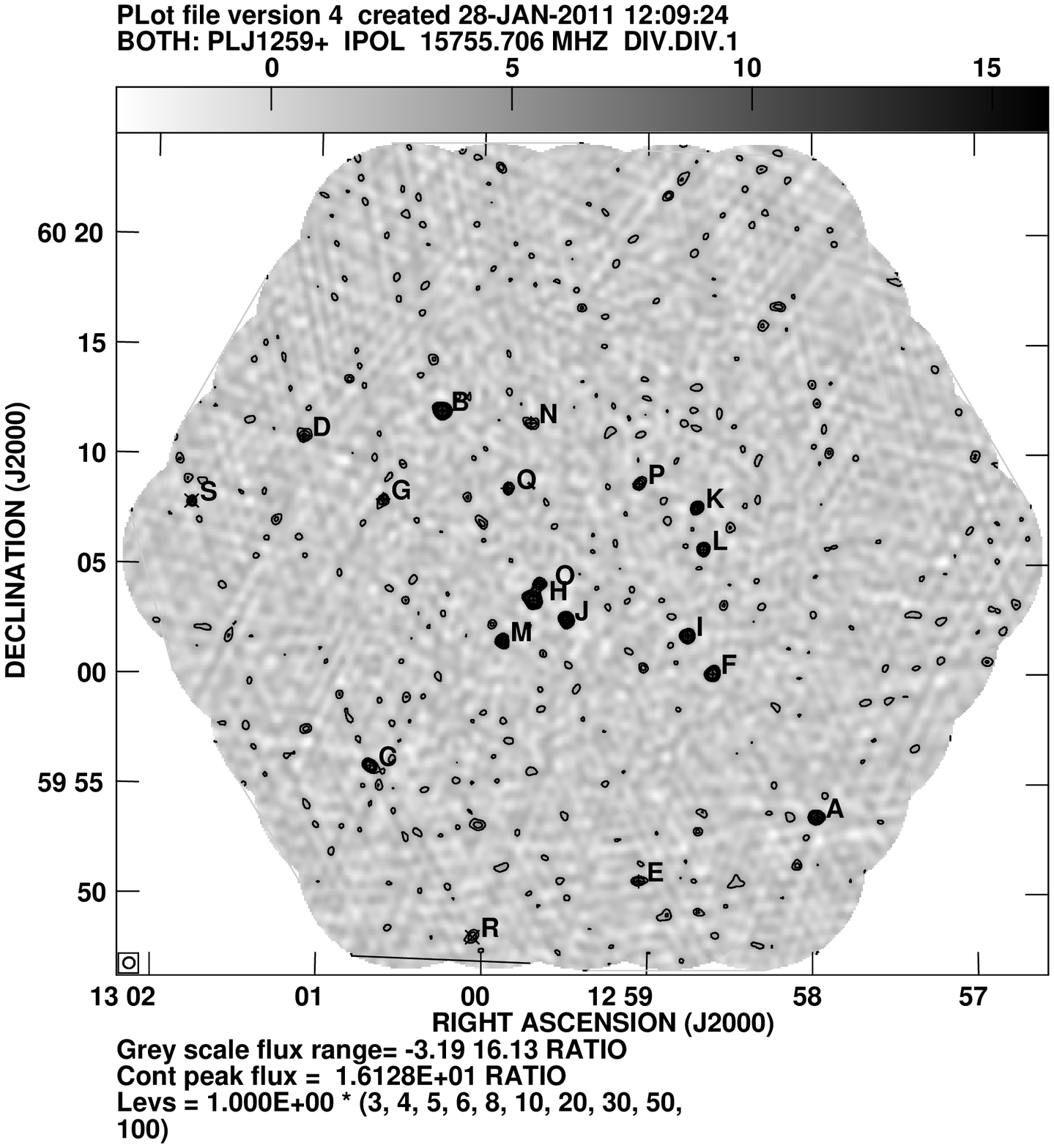}}
\caption{LA raster signal-to-noise maps of the two candidates,
with sources detected at $4\sigma$ and above marked; their properties can be found 
in Tables~\ref{tab:PLJ2226-sources}~and~\ref{tab:PLJ1259-sources}.
Contours are S/N$=1$, and follow levels $+3$, $+4$, $+5$, $+6$, $+8$, $+10$ and $+20$.
The small ellipse in the box in the lower-left shows the average synthesised beam.
\label{fig:LA-map}}
\end{center}
\end{figure}

To analyse the SA cluster observations we use a Bayesian analysis package, \textsc{McAdam}
(\citealt{marshall03} and \citealt{McAdam}). This uses \textsc{MultiNest}
(\citealt{feroz07} and \citealt{MultiNest}), an implementation of nested sampling
\citep{skilling04}, to efficiently explore the multidimensional parameter space and
to calculate the Bayesian evidence. This analysis has been applied to pointed observations
of known clusters (e.g. \citealt{bullet-like}; \citealt{the-eight-clusters}), 
and also to detect previously unknown clusters \citep{AMIblind}.
We use it here to model the effect of radio point sources on the SA data,
using the LA data as priors.
To transform the SA data into a format suitable for \textsc{McAdam}, the
$uv$-data for all good observations were concatenated together to make
a $uv$-\textsc{fits} file for each channel. These were then transformed into
lists of visibilities for the purpose of generating a covariance matrix for the
data. The covariance matrix describes the
terms which contribute to the data but are not part of the model, i.e.
the CMB, thermal noise from the telescope
and confusion noise from unresolved point sources. The latter were modelled using
the Tenth Cambridge Radio Survey (10C: \citealt{10C1}; \citealt{10C2}), integrating
the confusion power from zero up to the LA source detection limit.

The data were also binned into cells
of width $40\lambda$, determined in \citet{marshall-thesis}
as a suitable binning scale to reduce the size of the data to
a manageable level without adversely affecting the resulting inference of model parameters.
The binned data files were then analysed by \textsc{McAdam} using the
method described in \cite{the-eight-clusters}. Maps were made using \textsc{imagr}
in \textsc{aips} from each channel of the SA and LA observations;
however we here present only the combined-channel maps of the
SA and LA observations. The task \texttt{imean} was used on the LA individual maps to attach
the map noise to the map header, and \texttt{flatn} was then used to stitch the maps together, with
a primary beam correction applied using parameters shown in Tab.~\ref{tab:Fluxes-of-3C286}.

Source-finding was carried out on the LA data using software modified from that used in the
9C~survey \citep{waldram03}.
Spectral indices were fitted using LA maps of all six channels, assuming
source fluxes follow a power-law relation of $S\propto\nu^{-\alpha}$ for the AMI frequencies.
The properties of point sources detected at 4\,$\sigma$ and above by the LA
are used as priors when modelling the SA data in \textsc{McAdam}.
The position prior
used in \textsc{McAdam} is a delta function since the resolution of the LA
is around three times that of the SA; two exceptions are
discussed in Section~\ref{sec:PLJ2226}.  
LA flux densities generated the peak of Gaussian priors on the SA source flux densities,
with $\sigma$s set to the calibration error of five~per~cent of
the source flux density added in quadrature
with the local map noise. Spectral index ($\alpha$) priors were also set as Gaussians with $\sigma$ equal
to the error on the spectral index fit. These errors tend to be small ($\simeq0.4$)
for bright sources and large ($\simeq2$) for faint sources.
Fig.~\ref{fig:LA-map} shows the LA maps for the candidates with $4\sigma_\mathrm{LA}$
radio point sources marked; the positions
and properties of the sources are given in Tables~\ref{tab:PLJ2226-sources}~and~\ref{tab:PLJ1259-sources}.
\begin{table*}
\begin{center}
\tablefont
\begin{tabular}{cccccccccc}
\hline
ID & RA & Dec & $S_{\textrm{LA}}$/mJy & $\alpha_{\textrm{LA}}$  & $S_{\textsc{McAdam}}$/mJy & $\alpha_{\textsc{McAdam}}$\tabularnewline
\hline
A & 22 23 41.0 & +78 18 35.1 & $6.73\pm0.83$ & $0.55\pm0.21$    &  $8.71\pm0.09$ & $0.79\pm0.11$ \tabularnewline
B & 22 23 59.3 & +78 19 49.4 & $0.50\pm0.23$ & $0.48\pm1.55$    &  $0.50\pm0.09$ & $0.41\pm1.27$ \tabularnewline
\hline
C & 22 23 53.6 & +78 21 29.1 & $3.64\pm0.61$ & $0.43\pm0.45$	&  $4.32\pm0.08$ & $0.68\pm0.19$  \tabularnewline
D & 22 25 29.4 & +78 25 25.3 & $2.53\pm0.51$ & $0.89\pm0.53$	&  $2.47\pm0.08$ & $1.14\pm0.28$  \tabularnewline
E & 22 23 41.6 & +78 10 51.1 & $2.02\pm0.40$ & $1.14\pm1.58$	&  $2.18\pm0.12$ & $2.13\pm0.60$  \tabularnewline
F & 22 21 12.6 & +78 13 08.6 & $1.75\pm0.61$ & $-0.99\pm1.77$	&  $1.03\pm0.35$ & $-1.16\pm1.68$ \tabularnewline
G & 22 24 35.3 & +78 18 05.5 & $1.67\pm0.41$ & $2.13\pm0.61$	&  $1.80\pm0.07$ & $1.25\pm0.33$  \tabularnewline
H & 22 21 09.2 & +78 21 40.8 & $1.37\pm0.47$ & $0.37\pm1.73$    &  $1.63\pm0.23$ & $1.56\pm1.13$  \tabularnewline
I & 22 26 41.7 & +78 17 39.7 & $1.14\pm0.34$ & $0.74\pm0.70$	&  $1.16\pm0.09$ & $1.67\pm0.39$  \tabularnewline
J & 22 26 36.9 & +78 16 53.3 & $1.12\pm0.34$ & $0.28\pm0.72$	&  $1.35\pm0.09$ & $0.97\pm0.39$  \tabularnewline
K & 22 26 43.9 & +78 14 08.0 & $0.82\pm0.29$ & $0.63\pm1.01$	&  $0.74\pm0.07$ & $0.04\pm0.74$  \tabularnewline
L & 22 26 28.4 & +78 19 55.8 & $0.77\pm0.28$ & $-0.23\pm1.01$	&  $0.09\pm0.05$ & $-0.01\pm0.98$ \tabularnewline
M & 22 27 34.7 & +78 15 00.6 & $0.72\pm0.28$ & $0.25\pm1.04$	&  $0.91\pm0.07$ & $0.13\pm0.81$  \tabularnewline
N & 22 26 34.7 & +78 15 24.9 & $0.31\pm0.18$ & $1.15\pm1.56$	&  $0.41\pm0.06$ & $0.90\pm1.38$  \tabularnewline
O & 22 27 59.3 & +78 16 31.5 & $0.28\pm0.18$ & $0.90\pm1.70$    &  $0.16\pm0.06$ & $0.73\pm1.62$ \tabularnewline
\hline
P & 22 24 50.3 & +78 22 57.7 & $2.06\pm0.42$ & $0.30\pm0.73$	&  -- & -- \tabularnewline
Q & 22 26 49.9 & +78 24 28.7 & $0.53\pm0.27$ & $0.70\pm1.37$	&  -- & -- \tabularnewline
R & 22 23 56.9 & +78 15 12.8 & $0.47\pm0.23$ & $0.23\pm1.66$    &  -- & -- \tabularnewline
S & 22 30 06.4 & +78 14 34.4 & $0.40\pm0.40$ & $0.40\pm1.63$	&  -- & -- \tabularnewline
\hline
\end{tabular}
\caption{Sources found in the LA map of PLCKESZ\,G115.71+17.52, in order of modelling prior and LA flux density.
Sources~A~and~B were modelled with a Gaussian position prior of $\sigma=5$\arcsec\, to make allowance
for the extended emission detected in this area. Sources~P,~Q,~R,~and~S were directly subtracted from
the SA data to reduce the computational resources needed to model the sources in the centre.
\label{tab:PLJ2226-sources}}
\end{center}
\end{table*}
\begin{table*}
\begin{center}
\tablefont
\begin{tabular}{cccccccccc}
\hline
ID & RA & Dec & $S_{\textrm{LA}}$/mJy & $\alpha_{\textrm{LA}}$  & $S_{\textsc{McAdam}}$/mJy & $\alpha_{\textsc{McAdam}}$\tabularnewline
\hline
A & 12 57 58.3 & +59 53 32.9 & $3.97\pm0.79$ & $-1.08\pm1.35$	&  $2.91\pm0.37$  & $-0.49\pm1.15$ \tabularnewline
B & 13 00 15.4 & +60 12 01.3 & $2.66\pm0.55$ & $0.20\pm0.95$	&  $2.55\pm0.13$  & $-0.12\pm0.55$ \tabularnewline
C & 13 00 40.7 & +59 55 50.6 & $2.47\pm0.61$ & $0.22\pm1.46$	&  $3.26\pm0.24$  & $0.52\pm0.84$  \tabularnewline
D & 13 01 05.9 & +60 10 49.4 & $1.76\pm0.58$ & $-1.04\pm1.63$	&  $1.13\pm0.23$  & $-0.02\pm1.47$ \tabularnewline
E & 12 59 03.0 & +59 50 38.7 & $1.71\pm0.57$ & $-1.48\pm1.69$	&  $1.26\pm0.28$  & $-1.68\pm1.55$ \tabularnewline
F & 12 58 36.2 & +60 00 04.8 & $0.97\pm0.34$ & $0.64\pm1.39$	&  $0.69\pm0.10$  & $0.65\pm1.26$  \tabularnewline
G & 13 00 36.8 & +60 07 58.1 & $0.85\pm0.35$ & $-0.99\pm1.53$	&  $0.51\pm0.12$  & $-0.92\pm1.39$ \tabularnewline
H & 12 59 41.6 & +60 03 25.2 & $0.70\pm0.27$ & $-0.17\pm1.13$	&  $0.74\pm0.12$  & $1.27\pm0.75$  \tabularnewline
I & 12 58 45.3 & +60 01 49.0 & $0.68\pm0.27$ & $-1.54\pm1.37$	&  $0.58\pm0.09$  & $-1.92\pm1.22$ \tabularnewline
J & 12 59 29.6 & +60 02 31.8 & $0.64\pm0.26$ & $-0.77\pm1.12$	&  $0.63\pm0.09$  & $-0.25\pm0.95$ \tabularnewline
K & 12 58 42.0 & +60 07 39.5 & $0.47\pm0.23$ & $1.34\pm1.56$	&  $0.57\pm0.08$  & $1.67\pm1.27$  \tabularnewline
L & 12 58 39.6 & +60 05 45.0 & $0.46\pm0.23$ & $0.46\pm1.66$	&  $0.32\pm0.08$  & $0.66\pm1.48$  \tabularnewline
M & 12 59 52.9 & +60 01 33.2 & $0.41\pm0.21$ & $1.19\pm1.53$	&  $0.52\pm0.08$  & $0.89\pm1.38$  \tabularnewline
N & 12 59 42.9 & +60 11 29.1 & $0.38\pm0.21$ & $0.68\pm1.75$	&  $0.48\pm0.09$  & $0.09\pm1.47$  \tabularnewline
O & 12 59 39.6 & +60 04 08.9 & $0.34\pm0.19$ & $-1.22\pm1.64$	&  $0.34\pm0.09$  & $-0.76\pm1.47$ \tabularnewline
P & 12 59 03.1 & +60 08 43.8 & $0.29\pm0.18$ & $-0.79\pm1.71$	&  $0.19\pm0.07$  & $-0.51\pm1.64$ \tabularnewline
Q & 12 59 51.2 & +60 08 30.3 & $0.28\pm0.18$ & $-0.77\pm1.57$	&  $0.44\pm0.07$  & $0.41\pm1.33$  \tabularnewline
\hline
R & 13 00 03.2 & +59 48 03.9 & $4.68\pm1.35$ & $-0.47\pm1.77$	&   --            &   --           \tabularnewline
S & 13 01 46.8 & +60 07 49.5 & $3.66\pm1.04$ & $-0.74\pm1.72$	&   --            &   --           \tabularnewline
\hline
\end{tabular}
\caption{Sources found in the LA map of PLCKESZ\,G121.11+57.01, in order of modelling prior and LA flux density.
Sources~R~and~S were directly subtracted from
the SA data to reduce the computational resources needed to model the sources in the centre.
\label{tab:PLJ1259-sources}}
\end{center}
\end{table*}
\section{Results and Discussion}\label{sec:discussion}
After the source models were fitted to the SA data in \textsc{McAdam}, the sources
were subtracted from the data using the modelled source fluxes and spectral indices.
The data were then mapped again in \textsc{aips} to generate source-subtracted maps.
The parameters used for source-subtraction are the last two columns of
Tables~\ref{tab:PLJ2226-sources}~and~\ref{tab:PLJ1259-sources}.
\subsection{PLCKESZ\,G115.71+17.52}\label{sec:PLJ2226}
\begin{figure*}
\begin{center}
\subfloat[Before source subtraction\label{fig:PLJ2226_SA}]{\includegraphics[width=7.25cm,clip=]{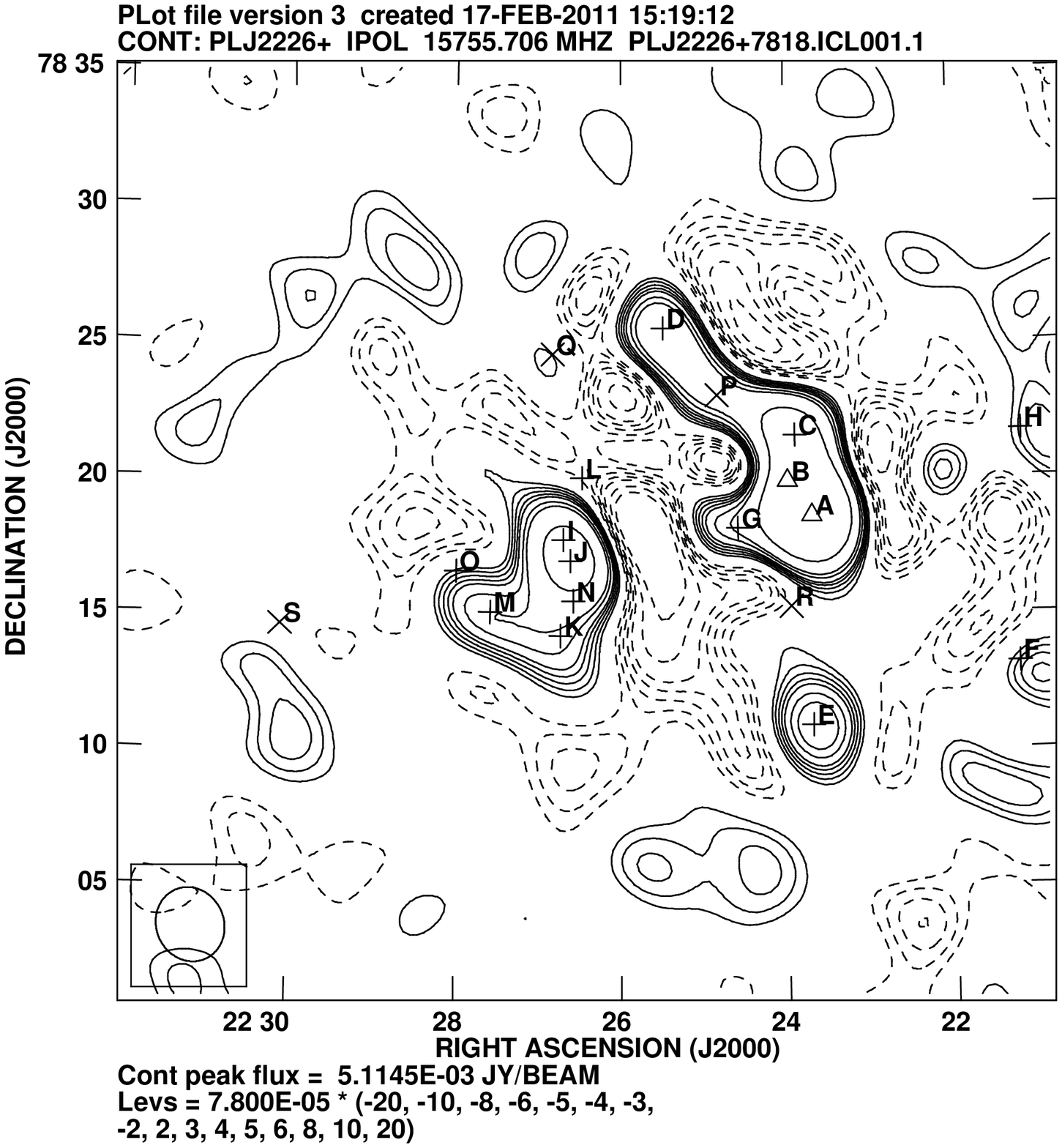}}\quad
\subfloat[After source subtraction\label{fig:PLJ2226_SS}]{\includegraphics[width=7.25cm,clip=]{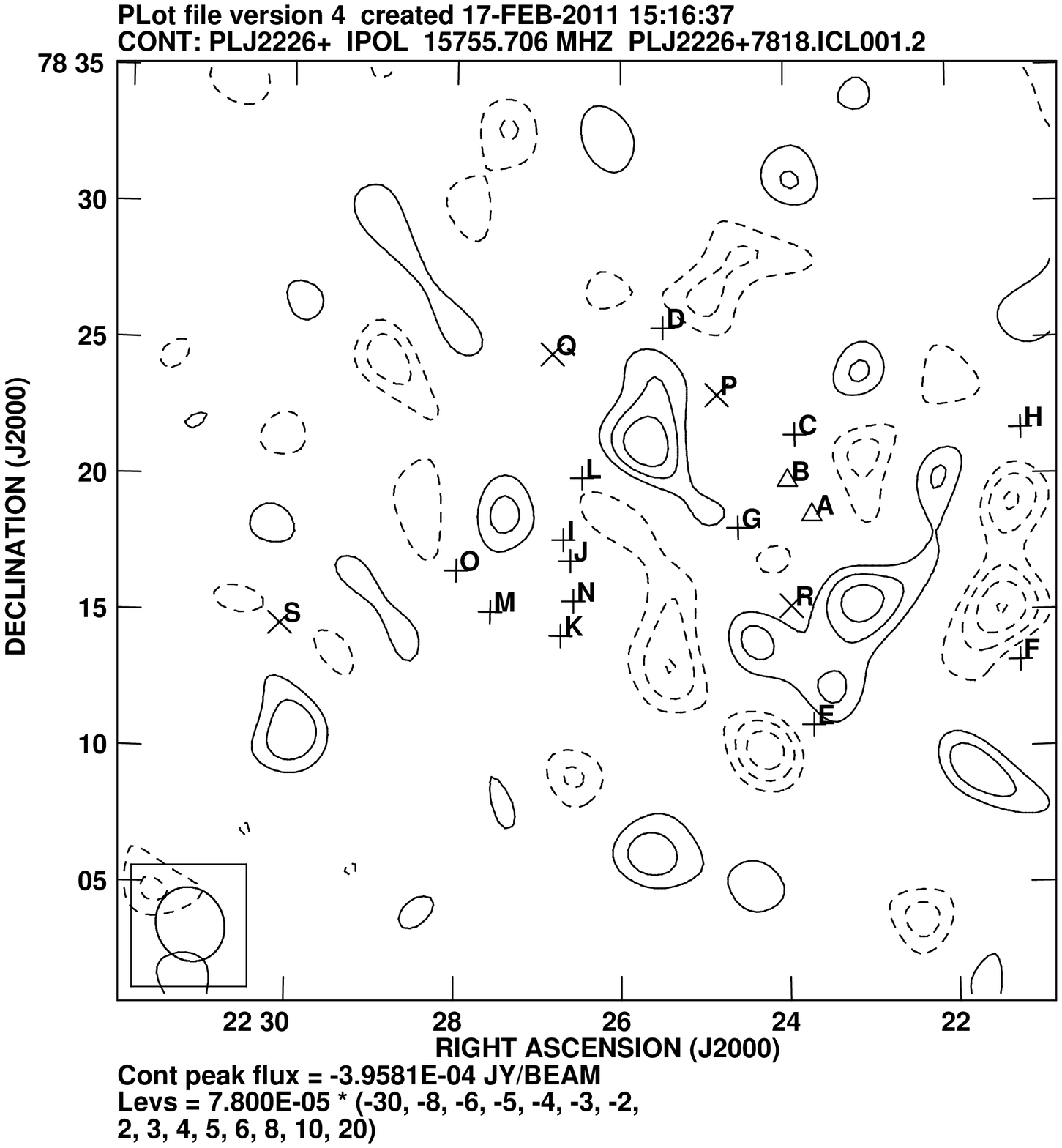}}
\caption{PLCKESZ\,G115.71+17.52 SA map before and after source subtraction; contours
are every $\sigma=78\mu$Jy, and follow levels
$-20$, $-10$, $-8$, $-6$, $-5$, $-4$, $-3$, $-2$, $+2$, $+3$, $+4$, $+5$, $+6$, $+8$, $+10$ and $+20$.
Sources marked with a `$+$' were modelled in \textsc{McAdam}, while those marked with
a `$\times$' were directly subtracted. The sources marked with a $\triangle$ were given a
Gaussian position priors of width 5\arcsec\, due to their slightly extended structure.
The parameters of the labelled sources can be found in Tab.~\ref{tab:PLJ2226-sources}.
The small ellipse in the box in the lower-left shows the synthesised beam.
\label{fig:PLJ2226-SA}}
\end{center}
\end{figure*}

The raw map for this candidate, Fig.~\ref{fig:PLJ2226_SA},
shows a very complicated region with both positive and negative
features. The sources appear to consist of two main complexes:
one to the south-east of the pointing centre, consisting of sources L, I, J, N, K, M and O.
To the west and north is a separate complex, consisting of sources D, P, C, B, G and the
extended source A. Given the extent of this source, and the uncertainty on whether source~B
was also extended, these were modelled with Gaussian position priors with $\sigma=5$\arcsec.
This is useful when dealing with extended emission as the location and magnitude of the
peak of such emission may differ
significantly when measured on different scales; it has previously been used to account for
small amounts of extended emission in \cite{AMIlocuss1} and \cite{AMIlocuss2}.

After source-subtraction (see Fig.~\ref{fig:PLJ2226_SS}),
extended positive residuals are visible directly to the
north of the pointing centre, and to the south of source~A.
These areas of emission have no obvious counterparts on the LA map (Fig.~\ref{fig:PLJ2226_LA})
and have SA peak flux densities $\simeq0.4$\,mJy.
However the complexes have integrated flux densities of $\simeq0.7$\,mJy each.
This seems to indicate that the emission is real, extended, and positive.

A difficulty arises when using an interferometer to measure structure of unknown
extent, as it is unclear how much will be resolved out due to the lack of
$uv$-coverage on large scales. This partially negates the natural filtering
advantage of using an interferometer for SZ detection, discussed in \cite{birkinshaw99}.
When deconvolving the dirty map with the synthesised beam
in the \textsc{clean} process \citep{clark-clean}, incomplete interferometric measurement
of extended structure can lead to artifacts, including negative signal. It is unclear whether
the negative signal 7\arcmin\, south-west of the pointing centre,
in the source-subtracted map, is purely artifactual from poor \textsc{clean}ing of the
extended positive emission, or also includes a component from SZ emission.
Publicly available \textit{ROSAT} data \citep{1999A+A...349..389V}
show elongated X-ray emission of $\simeq5$\arcmin extent
at the \textit{Planck} ESZ position. If this object is a galaxy cluster
then the SZ and X-ray emission are likely not to be strongly offset, so the south-west negative
feature (and other negative features on the map) are probably artifactual.

As to the source of the positive extended emission seen in this region, we note that
Galactic observations with the SA (see e.g. \citealt{2009MNRAS.396..365H}) tend to pick
up more extended emission closer to the Galactic plane. This is likely to be synchrotron
associated with regions which also
have bright thermal dust emission. The latter was also noted by \cite{PlanckESZ} as
a problem when measuring the SZ effect toward this object.
There is also the possibility that the two source complexes are extended
radio doubles from jet-producing galaxies, which would be more likely to
appear in a galaxy cluster than in the field.
\subsection{PLCKESZ\,G121.11+57.01}\label{sec:PLJ1259}
\begin{figure*}
\begin{center}
\subfloat[Before source subtraction.\label{fig:PLJ1259_SA}]{\includegraphics[width=7.25cm,clip=]{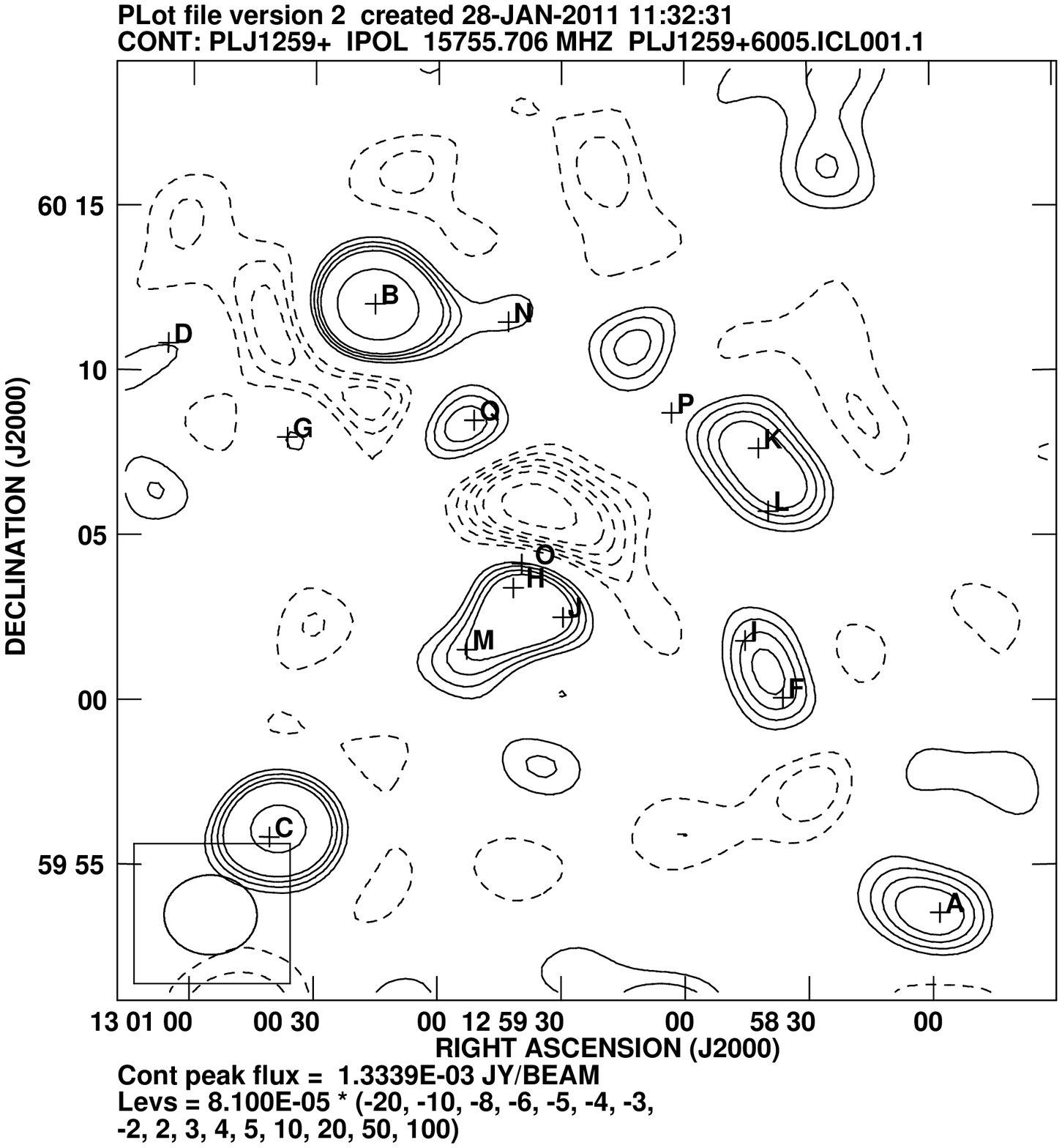}}\quad
\subfloat[After source subtraction.\label{fig:PLJ1259_SS}]{\includegraphics[width=7.25cm,clip=]{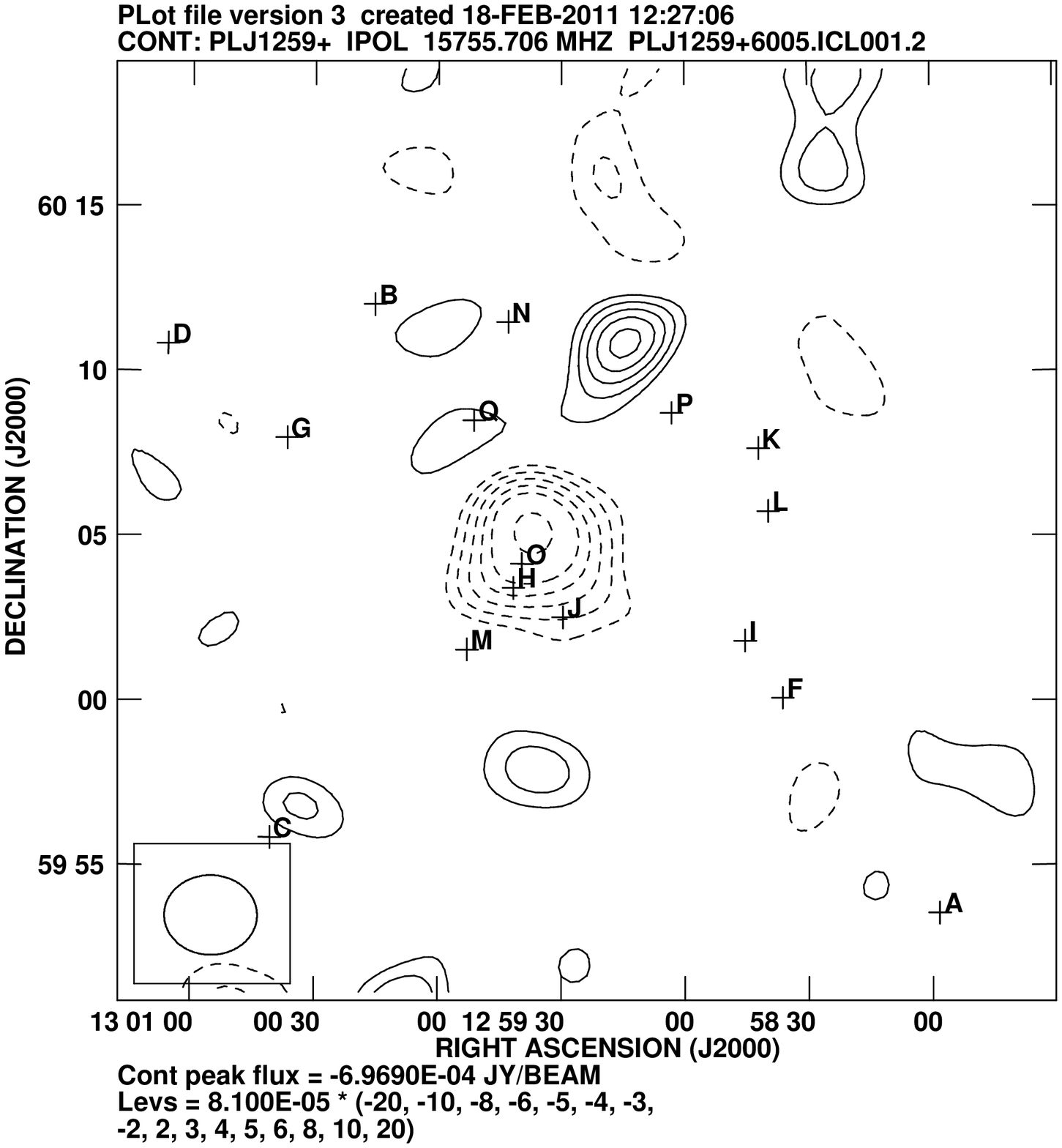}}
\caption{PLCKESZ\,G121.11+57.01 SA map before and after source subtraction; contours
are every $\sigma=81\mu$Jy, and follow the same levels as in Fig.\ref{fig:PLJ2226-SA};
annotations and markings are also the same. Sources E,~R, and~S are outside the field-of-view shown here,
and the latter two were directly subtracted from the SA data.
The parameters of the subtracted sources can be found in Tab.~\ref{tab:PLJ1259-sources}.
\label{fig:PLJ1259-SA}}
\end{center}
\end{figure*}

It is clear even before source subtraction that there is isolated excess negative signal in
the centre of the SA map, Fig.~\ref{fig:PLJ1259_SA}. After source-subtraction the
initially elliptical appearance is revealed as a close to circularly-symmetric SZ decrement,
Fig.~\ref{fig:PLJ1259_SS}. The 6$\sigma$ positive
residual between sources~P~and~N corresponds well to an overdensity of 3$\sigma$ features nearby in the LA map,
and is likely to be extended emission resolved-out by the LA. The feature is more than
one synthesised beam away from the cluster and appears alone, so should not affect measurement
of the cluster decrement.

From the source-subtracted map, we measure the integrated flux density
of the central negative signal as $-1.08\pm0.10$\,mJy.
We estimate the S/N of our measurement as 13, by dividing the
absolute value of the integrated flux density by the map noise of $0.081$mJy.
Thus we confirm the \textit{Planck} blind SZ detection of PLCKESZ\,G121.11+57.01 as a
galaxy cluster. We hope to investigate the full comparison of AMI and Planck data via common
cluster models in future publications.
% A future paper will discuss the full comparison of AMI and \textit{Planck}
%data via common cluster models.

We also measure a position offset from the ERCSC position, finding
%$\Delta_{\mathrm{RA}}=+1\farcm56\pm0\farcm21$ and $\Delta_{\mathrm{Dec}}=-0\farcm63\pm0\farcm31$.
$\Delta_{\mathrm{RA}}=+94\pm13$\arcsec\, and $\Delta_{\mathrm{Dec}}=-38\pm19$\arcsec,
resulting in an improved cluster position of
RA~12$^{\mathrm{h}}$~59$^{\mathrm{m}}$~36.4$^{\mathrm{s}}$, Dec~+60$^{\circ}$04\arcmin~46\farcs8.
\cite{2009ApJS..183..197W} identify a cluster of redshift 0.33 at
RA~12$^{\mathrm{h}}$~59$^{\mathrm{m}}$~33.4$^{\mathrm{s}}$, Dec~+60$^{\circ}$~04\arcmin~09\arcsec\,
in the Sloan Digital Sky Survey DR6 Galaxy Clusters Catalog. Given that the separation between
these two measurements is merely 44\arcsec, and the redshift is typical of clusters detected by \textit{Planck}
\citep{PlanckESZ}, it is likely that the two are the same object.
Publicly-available \textit{ROSAT} data \citep{1999A+A...349..389V} show only a few X-ray
photons, consistent with noise, so the cluster appears to have a low X-ray luminosity. 
\section{Conclusions}
We confirm the \textit{Planck} blind SZ detection of PLCKESZ\,G121.11+57.01 as a
galaxy cluster at $13\sigma$, with a 15\,GHz integrated flux density of 
 $-1.08\pm0.10$\,mJy.
Our observations improve the measurement of the cluster position to
RA~12$^{\mathrm{h}}$~59$^{\mathrm{m}}$~36.4$^{\mathrm{s}}$, Dec~+60$^{\circ}$~04\arcmin~46\farcs8.
We cannot positively identify PLCKESZ\,G115.71+17.52
as a galaxy cluster via SZ, as the data are consistent with both the presence of
an SZ decrement and extended positive emission, or simply the latter.
However we note that radio source environment may indicate the presence
of jets from active galaxies, often found in galaxy clusters.
\section{ACKNOWLEDGMENTS}
We thank the staff of the Mullard Radio Astronomy Observatory for their invaluable
assistance in the commissioning and operation of AMI, which
is supported by Cambridge University and the STFC.
Computational results were obtained using the COSMOS supercomputer
(DiRAC STFC HPC Facility).
This work was based on observations obtained with \textit{Planck}, an ESA science mission with
instruments and contributions directly funded by ESA Member States, NASA, and Canada.
This research made use of the \textit{ROSAT} Data Archive of the Max-Planck-Institut
f\"{u}r extraterrestrische Physik (MPE) at Garching, Germany.
MLD, TMOF, CRG, MO, MPS and TWS acknowledge the
support of PPARC/STFC studentships.
\bibliographystyle{philthesis}
\bibliography{references}
\label{lastpage}
\end{document}